\documentclass[11pt, oneside]{article} 
\usepackage[margin=2cm]{geometry}                		
\usepackage[parfill]{parskip}    		
\usepackage{graphicx}			
\usepackage[utf8]{inputenc}
\usepackage[english]{babel} 
\usepackage{amsmath,amssymb,amsthm,bm} 
\usepackage[color=yellow]{todonotes}
\usepackage{mathrsfs}
\usepackage{url}	
\usepackage{esint}
\usepackage[colorlinks]{hyperref}
\usepackage{enumerate}
\usepackage{outlines}[enumerate]
\usepackage{framed,comment,enumerate}
\usepackage{upgreek}
\usepackage{pgfplots}
\usepackage{tikz,tikz-cd}
\usetikzlibrary{positioning}
 \usepackage{float}
 \usepackage{rotating}
\usepackage{graphicx}
\usepackage{caption}
\usepackage{multicol}
\usepackage{cancel}
\usepackage{subcaption}
\usepackage[title]{appendix}
\usepackage{pgfgantt}
\usepackage{rotating}
\usepackage{tikzscale}
\extrafloats{100}
\numberwithin{equation}{section}

\theoremstyle{definition}

\DeclareFontFamily{U}{MnSymbolC}{}
\DeclareSymbolFont{MnSyC}{U}{MnSymbolC}{m}{n}
\DeclareFontShape{U}{MnSymbolC}{m}{n}{
    <-6>  MnSymbolC5
   <6-7>  MnSymbolC6
   <7-8>  MnSymbolC7
   <8-9>  MnSymbolC8
   <9-10> MnSymbolC9
  <10-12> MnSymbolC10
  <12->   MnSymbolC12}{}
\DeclareMathSymbol{\intprod}{\mathbin}{MnSyC}{'270}

\newcommand{\rem}[1]{}  

\newcommand{\scp}[2]{\left<#1\,,\,#2\right>}

\def\p{{\partial}}

\def\de{{\delta}}

\DeclareFontFamily{U}{mathx}{}
\DeclareFontShape{U}{mathx}{m}{n}{<-> mathx10}{}
\DeclareSymbolFont{mathx}{U}{mathx}{m}{n}
\DeclareMathAccent{\widehat}{0}{mathx}{"70}
\DeclareMathAccent{\widecheck}{0}{mathx}{"71}

\def\p{\partial}

\begin{document}

\title{\textbf{A Multiscale Camassa--Holm Equation}}
\author{Darryl D. Holm and Maneesh Kumar Singh  
\thanks{Department of Mathematics, Imperial College London}  
\footnote{Corresponding author: maneesh-kumar.singh@imperial.ac.uk} \\ 
\footnotesize
d.holm@ic.ac.uk, maneesh-kumar.singh@imperial.ac.uk 
\\  \small
}
\date{}

\maketitle

\begin{abstract}
A system of equations for Multiscale Geodesic Flow (MGF) is introduced whose solutions illustrate 
the paradigm of whorls within whorls within whorls, introduced by L. F. Richardson in 1922 to describe the 
cascade of energy in fluid turbulence. Numerical simulations are given for MGF on $S^1$, where the MGF 
equation comprises a multiscale generalisation of the Camassa-Holm equation (CHE) 
whose emergent singular solutions generalise the peakon solutions of the CHE.
\end{abstract}



\section{Introduction}
Based on the soliton-bearing Camassa--Holm (CH) equation on $S^1$ derived in \cite{camassa1993integrable} 
a new class of models for the mean motion 
of ideal incompressible fluids in three dimensions (3D) was proposed and analysed in \cite{holm-1998-euler}. 
In the 3D CH equation the Lagrangian--averaged amplitude of the fluctuations introduces a spatial scale, $\alpha$, 
and below this scale the Lagrangian transport velocity is filtered by nonlinear dispersion. 
This $\alpha$-filtering enhances the stability and regularity of the new fluid models without compromising either
their large scale behaviour, or their conservation laws. These models also describe geodesic motion on
the volume-preserving diffeomorphism group for a metric containing the $H^1$ norm of the fluid velocity.
The 3D CH equation came to be called the Lagrangian--Averaged Euler-$\alpha$ (LAE-$\alpha$) model
in \cite{marsden2000geometry} where the vector field solutions that generate the geodesic Lagrangian path of  
the 3D CH flow were proved to be smooth and converge to 
solutions of the Euler fluid equation in the limit $\alpha\to 0$.  

The \emph{viscous} 3D CH equation was proposed 
as a closure approximation for the Reynolds-averaged equations of the divergence-free 
Navier-Stokes flow in \cite{chen1998camassa}. 
This closure approximation was evaluated on turbulent channel and pipe flows with steady mean, and  
its analytical solutions for the mean velocity and the Reynolds
shear stress were found to be consistent with experiments in most of the flow region. 
In \cite{foias2002three}, the viscous 3D CH equation was shown to possess global strong solutions 
which converge to solutions of the Navier--Stokes fluid equation 
 in the limit $\alpha\to 0$.

Thereafter, the viscous 3D CH equation was called the Lagrangian-Averaged Navier--Stokes alpha 
(LANS-$\alpha$) model and was tested as a Large Eddy Simulation (LES) model in various
physical applications, including 
magnetohydrodynamics (LAMHD-$\alpha$) in astrophysics \cite{pietarila2006inertial},  as well as
rotation and stratification in oceanography \cite{hecht2008implementation}.
The LANS-$\alpha$ model also satisfies the K\'arm\'an--Howarth turbulence  theorem, \cite{holm2002karman}.

The present paper derives $N$-dimensional equations for \emph{multiscale} geodesic models of the
famous Richardson cascade of energy and circulation \cite{richardson1922weather}, as illustrated by computational simulations 
of their solutions on $S^1$.

\begin{figure}[H] 
    \centering
    \includegraphics[width=0.7\textwidth]{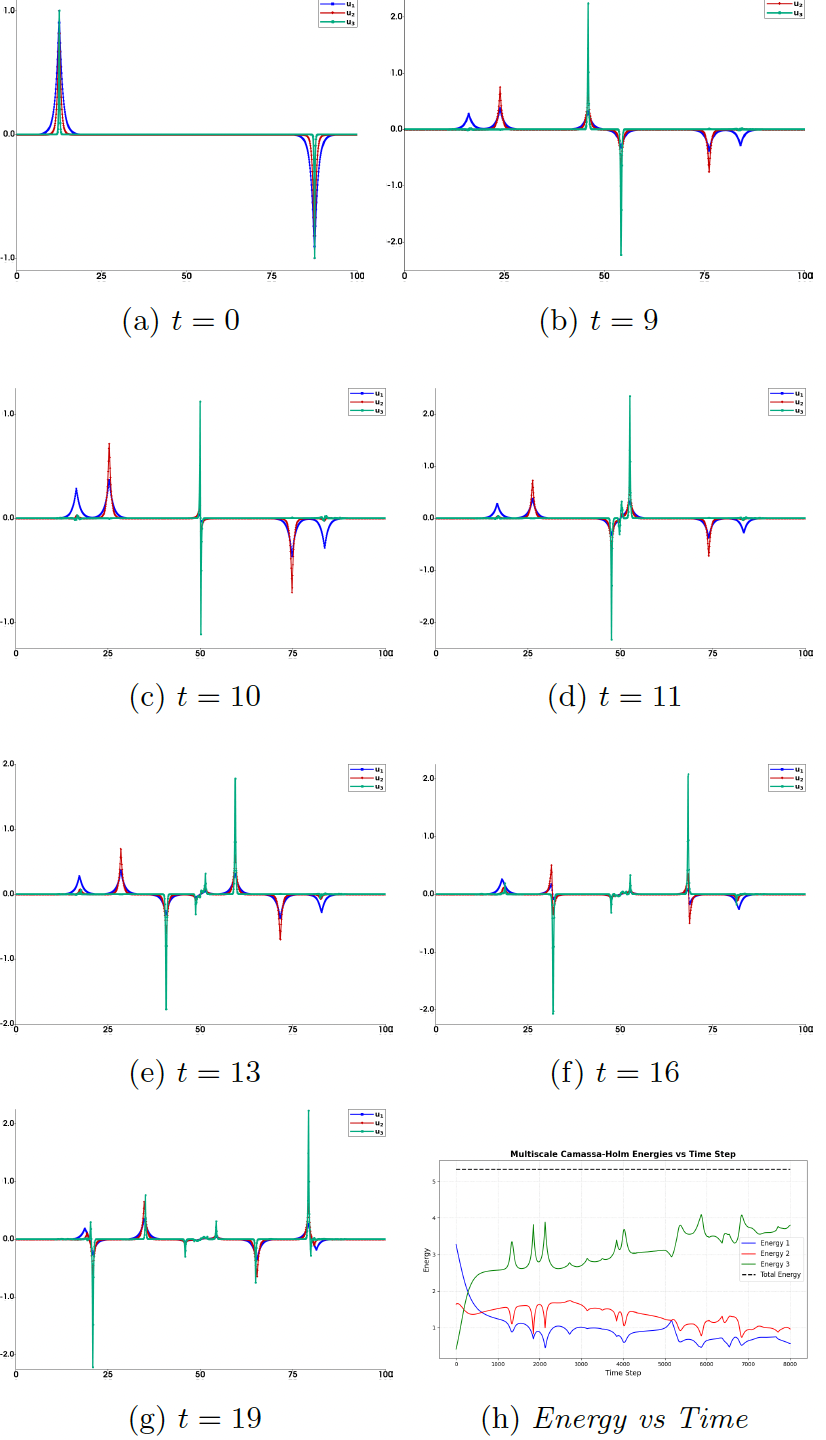}
    
    \caption{\label{erg-flo} \footnotesize The MGF equation \eqref{1D-mom-transport} is solved by midpoint scheme in time and Galerkin FEM in space. Fig. (a)-(g) show velocity profiles $u_1$ (blue), $u_2$ (red), $u_3$ (green). An initial 3-component peakon-antipeakon pair starts on either side of the periodic domain of length 100 at time $t=0$. A train forms of rightward moving (3,2,1) peakons and leftward moving (3,2,1) antipeakons at $t=9$. Head-on collisions for $t=10$ (3-3), $t=16$ (2-3) and $t=19$ (1-3). Fig. (h) shows the flow of total constant energy from 1-2-3 with sudden energy exchanges at the collision times.  }
    \label{fig:peakon_evolution} 
\end{figure}

\section{Multiscale geodesic flows (MGF)}
Geodesic flows on the manifold of smooth invertible maps 
$G$ acting on $\mathbb{R}^N$ may be lifted as curves $g_t\in G$ parametrised by time $t$ 
as $x_t = {g_t}x_0\in \mathbb{R}^N$. The flows $g_t\in G$ may be derived via Hamilton's principle 
$0 = \de S = \de \int \ell(u)\,dt$ for a kinetic energy 
Lagrangian $\ell(u)=\tfrac12 \|u\|^2 = \tfrac12\int_\mathbb{R} u\, Q^{op} u \,dx$ 
with vector-field velocity $u=\dot{g}_t{g_t}^{-1}\in \mathfrak{g}\simeq T_eG$ and positive definite, symmetric differential 
operator $Q^{op}$
\[
0 = \de S =  \de \int_0^T \ell(u) + \scp{m}{\dot{g}_t{g_t}^{-1} - u}_{\mathfrak{g}^*\times \mathfrak{g}}\,dt
\,.\]
Here, the brackets $\scp{\cdot}{\cdot}_{\mathfrak{g}^*\times \mathfrak{g}}$ denote $L^2$ pairing between 
the Lie algebra of vector fields $\mathfrak{g}$ and its dual $\mathfrak{g}^*$. 
The dual Lie algebra $\mathfrak{g}^*$ is a Poisson manifold, whose Lie--Poisson bracket is discussed below.

Variations of this Hamilton principle yield an evolution equation expressed by the \emph{Lie chain rule},
\[
\frac{d}{dt}{g_t}_*m(x,t)  = - \,{g_t}_*({\cal L}_{u} m)
\,,\quad\hbox{with}\quad u =  \dot{g}_t g_t^{-1}
\,,\]
in which ${g_t}_*$ denotes \emph{push-foward} by the map $g_t\in G$  \cite{holm1998euler,crisan2020variational}
Thus, the coordinate-free \emph{Lie derivative} ${\cal L}_{u}$ is defined as (minus) the tangent to the push-forward ${g_t}_*$
along the Eulerian vector field $u=\dot{g}_t g_t^{-1}\in T_eG$ which generates Lagrangian flow $g_t$ and satisfies the equation
\footnote{The Lie-derivative and differential-form coordinate-free notation used here is appropriate, since solutions of the CH equation follow geodesic paths on the manifold of smooth invertible maps (diffeomorphisms) with respect to the kinetic energy metric given by  $H^1(\mathbb{R}^N)$ for any integer $N$. Thus, for example, $m(x,t)={g_t}_{*}m(x_0,0)$ with $g_0=Id$ is the push-forward of the initial momentum density $m_0$ by the smooth invertible flow map $g_t$ depending on time $t$ for any number of dimensions, $N$.}
\[
\partial_t m  = -\, {\cal L}_{u} m = -\, (\partial_x m + m \partial_x)u  
\,,\]
with $m(x,t) = \de \ell/\de u = Q^{op} u$ for geodesic flow.  Inverting the operator $Q^{op}$
solves for $u =  \dot{g}_t g_t^{-1}$ from which $g_t$ may be reconstructed for 
Lagrangian fluid trajectories.

After Legendre transforming to the conserved energy Hamiltonian
$
h(m) = \scp{m}{u}_{\mathfrak{g}^*\times \mathfrak{g}} - \ell(u)
\,,$
the Lie-Poisson Hamiltonian formulation of geodesic flow on the Poisson manifold $m\in P=\mathfrak{g}^*$, 
with Lie-Poisson bracket $\{ f , h \}$ written in terms of Lie algebra bracket $ [u,v] =: {\rm ad}_uv$,
\begin{align*}
\{ f , h \} = - \scp{m} { [ \de f / \de m ,\de h / \de m ] }_{\mathfrak{g}^*\times \mathfrak{g}}
    = - \scp{ {\rm ad}^*_{\de h / \de m} m } {\de f / \de m }_{\mathfrak{g}^*\times \mathfrak{g}}
    = - \scp{ {\cal L}_{\de h / \de m} m } {\de f / \de m }_{\mathfrak{g}^*\times \mathfrak{g}}
\,,
\end{align*}
which will be extended here from GF to MGF by introducing a multiscale Hamiltonian.

\subsection{Lie--Poisson Hamiltonian MGF in 1D} 
The Hamiltonian for a 1D MGF system in $\mathbb{R}$ is 
\begin{align}
H(\{m\}) = \frac12\sum_{k=1}^n \int_\mathbb{R} m_{\alpha_k} K_{\alpha_k} * m_{\alpha_k}\,dx
\,,
\label{Ham-0}
\end{align}
with Eulerian fluid velocity $u_{\alpha_k}(x,t)$ in the sum over finer and finer discrete scales, 
$\alpha_k\in \mathbb{R}$ where $k=1,2,\dots,n,$ $\alpha_{k+1}/\alpha_k=2^{-k}$ and
$m_{\alpha_k}(x,t)$ is defined as 
\begin{align*}
m_{\alpha_k}(x,t) &= {Q^{op}}_{\alpha_k} u_{\alpha_k} := (1-\alpha_k^2\p_x^2)u_{\alpha_k}(x,t)
\,,\\
u_{\alpha_k} &= K_{\alpha_k} * m_{\alpha_k} 
=: \int_{\mathbb{R}} K_{\alpha_k}(|x-y|) m_{\alpha_k} (y,t)\,dy
\,,
\end{align*}
where $K_{\alpha_k}(|x-y|)$ is the Green function for the Helmholtz operator $(1-\alpha_k^2\p_x^2)$ on the real line, 
for boundary conditions $\lim_{|x|\to\infty}u_{\alpha_k}(x,t)=0$. 

The Lie--Poisson form of the dynamics for the starting Hamiltonian with $\delta h/\delta m_{\alpha_k}=u_{\alpha_k}$
and $n=3$ is \cite{holm2025geometric}
\begin{align*}
\p_t\begin{pmatrix} m_{\alpha_1} \\  \\ m_{\alpha_2} \\  \\ m_{\alpha_3} \end{pmatrix}
= -
	\begin{bmatrix} 
	\mathrm{ad}^*_{\Box} m_{\alpha_1} &  \mathrm{ad}^*_{\Box} m_{\alpha_2}  & \mathrm{ad}^*_{\Box} m_{\alpha_3}
	\\ \\
	\mathrm{ad}^*_{\Box} m_{\alpha_2} & \mathrm{ad}^*_{\Box} m_{\alpha_2} & \mathrm{ad}^*_{\Box} m_{\alpha_3}
	\\ \\
	\mathrm{ad}^*_{\Box} m_{\alpha_3} & \mathrm{ad}^*_{\Box} m_{\alpha_3} & \mathrm{ad}^*_{\Box} m_{\alpha_3}
	\end{bmatrix} 
  \begin{pmatrix}
    u_{\alpha_1} \\ \\
    u_{\alpha_2} \\ \\
    u_{\alpha_3}
    \end{pmatrix}
    = -
    \begin{pmatrix} \mathrm{ad}^*_{u_{\alpha_1}} m_{\alpha_1} +  \mathrm{ad}^*_{u_{\alpha_2}} m_{\alpha_2} +  \mathrm{ad}^*_{u_{\alpha_3}} m_{\alpha_3} 
    \\ \\
    \mathrm{ad}^*_{u_{\alpha_1} + u_{\alpha_2}} m_{\alpha_2} +  \mathrm{ad}^*_{u_{\alpha_3}} m_{\alpha_3} 
    \\ \\
    \mathrm{ad}^*_{u_{\alpha_1} + u_{\alpha_2} + u_{\alpha_3}} m_{\alpha_3}
    \end{pmatrix},
\end{align*}
in which the higher (smaller) scales are seen to apply forces to the lower (larger) scales, while the sum of velocities of the lower (larger) 
scales transports the momenta of the higher (smaller) scales.

The matrix operator in square brackets above defines the following semidirect-product Lie-Poisson bracket \cite{holm1998euler}
\begin{align}
\begin{split}
\frac{df}{dt} = \big\{f,h\big\} &= \sum_{k=1}^n\scp{ m_{\alpha_k}} {\left[ \frac{\delta f}{\delta m_{\alpha_k}} , \frac{\delta h}{\delta m_{\alpha_k}}\right] }
 = - \sum_{k=1}^n \scp{ m_{\alpha_k} } { {\rm ad}_{ \frac{\delta h}{\delta m_{\alpha_k} }  }  \frac{\delta f}{\delta m_{\alpha_k}}  }
\\&= - \sum_{k=1}^n \scp{  {\rm ad}^*_{ \frac{\delta h}{\delta m_{\alpha_k} }  }  m_{\alpha_k} } { \frac{\delta f}{\delta m_{\alpha_k}}  }
,\end{split}
\label{LiePoisson-Brkt}
\end{align}
where $\scp{\cdot}{\cdot}$ denotes $L^2$ pairing of functions and  $[\cdot, \cdot]$ denotes the Lie algebra bracket of vector fields. That is,
\[
\bigg[\frac{\delta f}{\delta m}
, \frac{\delta h}{\delta m}\bigg]
=
{\rm ad}_\frac{\delta h}{\delta m}\frac{\delta f}{\delta m}
=
\frac{\delta f}{\delta m}\partial_x\frac{\delta h}{\delta m}
-
\frac{\delta h}{\delta m}\partial_x\frac{\delta f}{\delta m}
.
\]
on the dual  of the following \emph{nested} semidirect product Lie algebra 
$
\mathfrak{s} = \mathfrak{g}_{\alpha_1}\ \circledS  \big(\mathfrak{g}_{\alpha_2}\ \circledS\ \mathfrak{g}_{\alpha_3} \ \big)
$,
with dual coordinates given by $m_{\alpha_k}\in \mathfrak{g}_{\alpha_k}^*$ dual to $u_{\alpha_k}\in \mathfrak{g}_{\alpha_k}$ 
correspondingly for $k=1,2,3$, and symbol $\circledS$ denotes the semidirect product operation \cite{holm1998euler}.

Physically, the Lie-Poisson bracket $\{f,h\}$ in \eqref{LiePoisson-Brkt} refers to three types of nested fluid flow.
Its decreasing filter widths $\alpha_k$ for $\alpha_{k+1}/\alpha_k=2^{-k}$ mimic L. F. Richardson's famous ``whorls within whorls within whorls'' reference in characterising fluid dynamics \cite{richardson1922weather}.
\begin{quote}
\textit{Big whorls have little whorls, that feed on their velocity, \\and little whorls have lesser whorls, and so on to viscosity.}

-- L. F. Richardson (1922)
\end{quote}
In a mathematical sense, then, the Lie--Poisson dynamics $\frac{df}{dt}=\{f,h\}$ in \eqref{LiePoisson-Brkt} mimics `Richardson's triple'.
Namely, Richardson's triple is modelled as the composition of three non-commutative right group actions of the smooth invertible maps 
$G_{\alpha_1}\times G_{\alpha_2}\times G_{\alpha_3}$. the corresponding fluid velocities are
\begin{align*}
u_{\alpha_1} &\in \mathfrak{g}_{\alpha_1} \otimes u_{\alpha_2} \in \mathfrak{g}_{\alpha_2}/G_{\alpha_1}
\otimes u_{\alpha_3} \in  (\mathfrak{g}_{\alpha_3}/G_{\alpha_2})/G_{\alpha_1}
\,.\end{align*}
Here, the flow of $G_{\alpha_1}$ can represent the big whorls,  the flow of $G_{\alpha_2}$ can represent the little whorls carried in the frame of motion of  the big whorls, and the flow of $G_{\alpha_3}$ can represent the lesser whorls carried along successively by the two other whorls. 

\subsection{Diagonalising the Hamiltonian formulation} 

The linear transformation of variables to 
\[
\mu_{\alpha_1}= m_{\alpha_1} -m_{\alpha_2}\,,\quad\mu_{\alpha_2}= m_{\alpha_2} -m_{\alpha_3}\,,\quad \mu_{\alpha_3} = m_{\alpha_3}\,,
\]
diagonalises the previous \textit{entangled} Lie--Poisson matrix 
 and leads to equivalent untangled equations, 
\begin{align*}
\p_t\begin{pmatrix} \mu_{\alpha_1}  \\ \mu_{\alpha_2}  \\ \mu_{\alpha_3} \end{pmatrix}
%
= -
    \begin{pmatrix} \mathrm{ad}^*_{u_{\alpha_1}} \mu_{\alpha_1} 
    \\
    \mathrm{ad}^*_{(u_{\alpha_1}+u_{\alpha_2})} \mu_{\alpha_2}
    \\
    \mathrm{ad}^*_{(u_{\alpha_1}+u_{\alpha_2}+u_{\alpha_3})} \mu_{\alpha_3}    \end{pmatrix} 
= -
    \begin{pmatrix}  \big(\p_x \mu_{\alpha_1} + \mu_{\alpha_1}\p_x \big)u_{\alpha_1}
    \\
    \big(\p_x \mu_{\alpha_2} + \mu_{\alpha_2}\p_x\big)(u_{\alpha_1}+u_{\alpha_2})
    \\
    \big(\p_x \mu_{\alpha_3} + \mu_{\alpha_3}\p_x\big)(u_{\alpha_1}+u_{\alpha_2}+u_{\alpha_3}) 
    \end{pmatrix} 
\,,
\end{align*}
where $\p_x$ denotes partial derivative along the real line, $\mathbb{R}$ and boundary conditions are chosen so that $\lim_{|x|\to\infty}u_{\alpha_k}(x,t)=0$ for $k=1,2,3$.
The untangled equations arise from a diagonal Poisson bracket $\{f,h\}=\langle\mu,[df,dh]\rangle$ on the dual of the following direct product Lie algebra
$\mathfrak{s}_{diag} = \mathfrak{g}_{\alpha_1}\ \otimes  \mathfrak{g}_{\alpha_2}\ \otimes\ \mathfrak{g}_{\alpha_3} $.

Untangled equations for momentum differences is typical of Lagrangian reduction by stages 
based on the composition of Lie group actions \cite{holm1998euler}.
Consequently, the sum of velocities of larger scales transports the pairwise differences of momentum densities of the smaller scales, 
\begin{equation}
\p_t \mu_{\alpha_k} + {\rm ad}^*_{\sum_{j=1}^k\! u_{\alpha_j} }\mu_{\alpha_k}   = 0
\,.
\label{1D-mom-transport}
\end{equation}

%
%

\subsection{Multiscale measure-valued peakon dynamics} 

Remarkably, the form of the dynamics 
in terms of pairwise differences in singular momentum densities at smaller scales 
admits emergent measure-valued solutions. Namely, the singular solution for  $N_i$ \emph{peakons},
\begin{equation*}
\mu_{\alpha_i} (x,t) = \sum_{a=1}^{N_i}\, 
p^{(a)}_i (t)\,\delta\big(x - x^{(a)}_i(t)\big) 
\,,
\end{equation*}
in which $x\in\mathbb{R}$ and the parameters $p^{(a)}_i$ are the momenta of the $N_i$ peakons at the $i$-th level. 

The starting Hamiltonian in \eqref{Ham-0} may be written as
\[
H = \frac12 \sum_{k=1}^n \int_\mathbb{R}  u_{\alpha_k} m_{\alpha_k} \,dx
= \frac12 \sum_{k=1}^n \int_\mathbb{R} 
u_{\alpha_k} \sum_{j=k}^n \,\mu_{\alpha_j}\,dx
\]
with velocity $u_{\alpha_k}$ written in terms of pairwise differences in singular momentum density, given by
\begin{equation*}
u_{\alpha_k}(x,t) =  \sum_{i=k}^n K_{\alpha_i} * \mu_{\alpha_i}
= \sum_{i=k}^n\sum_{b=1}^{N_i}\, 
p^{(b)}_i(t)\, K_{\alpha_i} ({x}-{x}^{(b)}_i(t))
.
\end{equation*}
This expression reveals the canonical Hamiltonian,
\begin{align*}
H (\{x\},\{p\})
=
 \frac12 \sum_{i<j=1}^{n}  
 \sum_{a=1}^{N_j}  \sum_{b=1}^{N_i} p^{(a)}_ j \, p^{(b)}_i
\left[K_{\alpha_i} \left( \big|{x}_j^{(a)}-{x}^{(b)}_i \big|\right)
+
K_{\alpha_j} \left( \big|{x}_i^{(a)}-{x}^{(b)}_j \big| \right)\right]
\end{align*}
with Green function kernel for the Helmholtz operator $(1-\alpha_k^2\Delta)$ given by
$K_{\alpha_k}(r) := \frac{1}{2} e^{-r/\alpha_k}$ for $S^1$ where the canonically conjugate fields are $(\{x\},\{p\})$.
For details of the corresponding emergent measure-valued solutions for the $N$-dimensional  Camassa--Holm
equation, see \cite{holm2005momentum}.

\subsection{3D Multiscale Euler Flow} 
In the 3D multiscale Euler system, the role of $\mu_{\alpha_k}$ is played by the vorticity difference  
$\varpi_{\alpha_k} := \omega_{\alpha_k} -  \omega_{\alpha_{k+1}}$ with $\omega_{\alpha_k}:={\rm curl}Q^{op}u _{\alpha_k} $ 
and ${\rm div} u_{\alpha_k} = 0$. Thus, the previous equation for the general case becomes for the 3D multiscale Euler system,
\begin{equation*}
\p_t \varpi_{\alpha_k} - {\rm curl} \Big({\sum_{j=1}^k u_{\alpha_j} \times \varpi_{\alpha_k}\Big)   = 0
\,,} 
\end{equation*}
and one finds the multiscale Euler Kelvin theorem,
\begin{equation*}
\frac{d}{dt} \oint_{C\big({\sum_{j=1}^k u_{\alpha_j} \big)}} \varpi_{\alpha_k} \cdot dS = 0
\,.
\end{equation*}
The multiscale $\varpi_{\alpha_k}$ dynamical system follows from the conserved kinetic energy Hamiltonian
\begin{equation*}
H(\varpi_{\alpha_j})=\frac12 \sum_{k=1}^n \int 
\psi_{\alpha_k} \cdot \sum_{j=k}^n\varpi_{\alpha_j}\,{\rm d}V
\,,\label{velocity-Ham}
\end{equation*}
with divergence-free vector stream function ${\psi}_{\alpha_k}={\rm curl}^{-1}{u}_{\alpha_k}$,
for the following Lie--Poisson bracket for smooth real functionals $(F,\,H)$ of $\varpi_{\alpha_k}$,
\begin{align*}
\begin{split}
\{F,\,H\}(\varpi_{\alpha_k} )
&=
\int \varpi_{\alpha_k} \cdot 
\Big( \,{\rm curl}\, \frac{\delta F}{\delta \varpi_{\alpha_k}} 
\times   \,{\rm curl}\,\frac{\delta H}{\delta \varpi_{\alpha_k}}   \Big)
{\rm d}V
\\
\frac{dF}{dt} =
&- \int \frac{\delta F}{\delta \varpi_{\alpha_k} }  \cdot \,{\rm curl}\,
\Big( \varpi_{\alpha_k} \times  \,{\rm curl}\, \frac{\delta H}{\delta \varpi_{\alpha_k}}   \Big)
{\rm d}V
\end{split}
\label{3D-vorticity-brkt}
\end{align*}
for which
${\rm curl}\,( {\delta H}/{\delta \varpi_{\alpha_k}} ) = \sum_{k=1}^m{u}_{\alpha_k}$
and the $\varpi_{\alpha_k}$ equation follows.

The helicity of $\varpi_{\alpha_k}$, defined for $k=1,2,\dots,n$ by
\[
\Lambda_{\alpha_k} := \int
 \omega_{\alpha_k}\cdot {\rm curl}^{-1}\omega_{\alpha_k}\,{\rm d}V
\]
is conserved, since $\frac{d}{dt}\Lambda_{\alpha_k} = \{\Lambda_{\alpha_k},\,H\}(\varpi_{\alpha_k} )=0$.


\section{Conclusion and outlook}
This note has derived $N$-dimensional equations for \emph{multiscale} geodesic models of the
famous Richardson cascade of energy and circulation \cite{richardson1922weather},
whose emergent singular solution properties are illustrated in $S^1$ by the computational simulations in Fig. \ref{erg-flo}.

A plethora of new opportunities remain for modelling and analysing multiscale phenomena.
For example, 
\begin{itemize}

\item 
One may include effects of advected quantities such as compressibility, heat transfer, or magnetic fields, as in  \cite{holm1998euler,holm2009geodesic},
in which case equation \eqref{1D-mom-transport} will no longer be geodesic. Instead, the equation set will enlarge to include 
transport relations and forces due to advected inertial, thermodynamic, magnetic, or electrodynamic multiscale quantities. Such an equation set could 
represent multiscale effects in plasma physics, for example, or even advected multiscale stochastic fluctuations.


\item 
It has not escaped our notice that the equation \eqref{1D-mom-transport} derived here for the Richardson cascade of energy shown in Fig. \ref{erg-flo} may potentially be associated with multifractal models of turbulence, as discussed in \cite{argoul1989wavelet,zhou2022wavelet}.
Although this topic is beyond the scope of the present paper, one may consider a stochastic transport perturbation in Richardson's triple as in \cite{holm2019stochastic}, to be studied via wavelet analysis. 
\end{itemize}

\noindent
{\bf Acknowledgements.}
    We are grateful to B. Chapron, C. Cotter, J. D. Gibbon, C. Tronci, and J. Woodfield for their thoughtful suggestions during the course of this work which have improved or clarified the interpretation of its results. DH was partially supported during the present work by Office of Naval Research (ONR) grant award N00014-22-1-2082, Stochastic Parameterisation of Ocean Turbulence for Observational Networks. DH and MKS were  partially supported  by European Research Council (ERC) Synergy grant Stochastic Transport in Upper Ocean Dynamics (STUOD) -- DLV-856408.

\bibliographystyle{apalike}
\bibliography{references}

\end{document}